\documentclass[aps, 10pt, prd, twocolumn, showpacs, longbibliography]{revtex4-2}

\usepackage{amsmath}
\usepackage{amssymb}
\usepackage{graphicx}
\usepackage{bm}
\usepackage{lmodern}
\usepackage{verbatim}
\usepackage{units}
\usepackage{graphicx}
\usepackage[utf8]{inputenc}
\usepackage{xcolor}

\usepackage[english]{babel}

\definecolor{darkgreen}{rgb}{0.0, 0.5, 0.0}

\begin{document}

\title{Gravitational Repulsion in an Expanding Ball of Dust} 
\author{Diogo P. L. Bragança}
\email{diogobraganca@alumni.stanford.edu}
\affiliation{Kavli Institute for Particle Astrophysics and Cosmology - KIPAC, Department of Physics, Stanford University, Stanford, CA 94305, USA}

\begin{abstract}
In general relativity, there is a velocity dependent term in the gravitational acceleration of a test particle for an observer at infinity. 
Depending on the direction of motion and the speed, that term can be repulsive. 
We show that this is also the case in the Parametrized Post-Newtonian (PPN) formalism.
We compute the magnitude of that repulsive term for an expanding sphere of dust observed at infinity, and find that it could mimic the effect of a cosmological constant.
The time evolution of such an expanding ball of dust for an observer at infinity is calculated, and compared with the standard $\Lambda$CDM model. 
We find that the so-called coincidence problem does not exist for such a model as the energy density attributed to the expansion is always of the same order as the matter energy density.
\end{abstract}

\maketitle

\tableofcontents

\section{Introduction}

Since Milne and McCrea~\cite{Milne1934,McCrea1934,McCrea1955} found that the Friedmann equations could be derived from Newtonian mechanics, the so-called Newtonian cosmology consisting in the study of an expanding ball of fluid has consistently been explored and interpreted, especially in comparison with relativistic cosmology. 
Harrison~\cite{Harrison1965} used Newtonian gravity to derive the Friedmann equations from a partitioned Universe.
Callan, Dicke, and Peebles~\cite{Callan1965} also confirmed that a Newtonian mechanics is sufficient to describe the real expanding universe (and not just a crude classical model).
Tipler~\cite{Tipler1996,Tipler1996a} even showed that Newtonian cosmology could be made as rigorous as relativistic cosmology provided that Newtonian gravity is formulated in geometrical language, proving also that Newtonian cosmology is more general that Friedmannian cosmology.
Reis~\cite{Reis2003} showed that even if the background evolution in Newtonian and relativistic cosmology could match, the growth of perturbations in the two regimes is not exactly the same when pressure is non-zero. 
Ellis and Gibbons~\cite{Ellis2014} built a discrete Newtonian cosmology from the bottom up using only point-particles, again finding the same equations as relativistic cosmology.

Newtonian cosmology has also inspired work in describing an expanding ball of fluid from a fully relativistic perspective. 
In that regard, Bili\'{c} and Toli\'{c}~\cite{Bilic2013} showed that this system can reproduce any FLRW cosmology by choosing an appropriate potential.
More recently, Faraoni and Atieh~\cite{Faraoni2020} studied a similar system, studying quasi-local energy aspects and showing that there is a symmetry correspondence between the Einstein-Friedmann equations and the Newtonian system.

In parallel to this study of Newtonian cosmology, it has also been observed that general relativity contains an effectively repulsive contribution to the gravitational force.
Hilbert~\cite{Hilbert1917} was the first to make this observation in a Schwarzschild spacetime for an observer at infinity. 
McVittie~\cite{McVittie1965} also considered such repulsive contributions.
Zel'dovich and Novikov~\cite{Zeldovich1965} claimed that this was merely a coordinate effect (a falling object seems to slow down at the event horizon).
McGruder~\cite{McGruder1982} provided a review of this particular discussion in the twentieth century and noticed as Hilbert that if the particle velocity was great enough anywhere in Schwarzschild spacetime then an observer at infinity would always observe a repulsive force.
More recently, Gorkavyi and Vasilkov~\cite{Gorkavyi2016} claimed that there could also be a repulsive force in general relativity if the gravitational mass of the system decreases (for example by gravitational radiation).
McGruder~\cite{McGruder2017} even hypothesized that the gravitational repulsion in Schwarzschild spacetime could explain cosmic rays.
Gr\o{}n~\cite{Gron2018} showed that if the velocity of the particle is zero, then there is never a gravitational repulsion in the Schwarzschild field.

Given these two research topics (Newtonian cosmology and gravitational repulsion), it then becomes interesting to combine them and analyze the relativistic gravitational repulsion phenomenon in an expanding ball of dust (the system for Newtonian cosmology).
An important motivation for this is that it could possibly mimic a cosmological constant type repulsion in a Newtonian cosmology.
In fact, there has been debate in the literature as to why the cosmological constant value is so small (compared to a na\"{i}ve computation of the quantum vacuum energy density) and thus un-natural, and also why the energy density associated with the cosmological constant is of the same order as the energy density associated with matter today (when it was not so in the past and will not be so in the arbitrary future). 
This latter observation is called the cosmic ``coincidence problem'' (a review of this issue is provided in~\cite{Velten2014}).

In this work, we analyze the dynamics of an expanding ball of dust for an observer unaffected by gravity (i.e. at infinity). 
We will find that the gravitational repulsion indeed has a significant effect, especially when the velocity is greater than $c/\sqrt{3}$, locally mimicking a cosmological constant.
We also show that there is no coincidence problem for an observer at infinity if the accelerated expansion is purely due to general relativity.

This paper is organized as follows. 
In Sec.~\ref{sec:repulsion}, we give a recap on situations where an observer may detect gravitational repulsion in general relativity. 
In particular, we check that an observer at infinity may detect a repulsion  in the Schwarzschild spacetime and in a two-body situation using a post-Newtonian formalism.  
In Sec.~\ref{sec:ball}, we mathematically construct an expanding ball of dust in general relativity using the Oppenheimer-Snyder metric~\cite{Oppenheimer1939}, and we check that the gravitational repulsion in this situation can mimic a cosmological constant for an observer located at infinity.
In Sec.~\ref{sec:cosmology}, we study the effective Friedman equation describing the scale factor's evolution and compare it to $\Lambda$CDM cosmology. In particular, it is shown that for the same current cosmological parameters the Universe is older in this framework than in $\Lambda$CDM, and that there is no ``coincidence problem''.
In Sec.~\ref{sec:conclusion}, we discuss the result and conclude.

\section{Gravitational repulsion in Schwarzschild and in the PPN formalism} 
\label{sec:repulsion}

Within the context of general relativity, we can find a so-called gravitational repulsion in different setups. 
In this section, we show how to get these terms in Schwarzschild spacetime and in the Parametrized Post-Newtonian (PPN) formalism.

\subsection{Repulsion in Schwarzschild spacetime}

Let us briefly review how we can obtain a gravitational repulsion in a Schwarzschild spacetime.
In this section we follow some calculations presented in~\cite{Hilbert1917,McGruder1982}.
In Schwarzschild coordinates, the metric is given by 
\begin{equation}
\begin{split}
	ds^2 = -\left(1-\frac{2m}{r}\right)c^2 dt^2 +&  \frac{dr^2}{1-\frac{2m}{r}}\\
&	\quad + r^2\left(d\theta^2 + \sin^2\theta\,d\phi^2\right)\,,
\end{split}
\end{equation}
where $m = GM/c^2$ is the geometrical mass, $G$ is Newton's gravitational constant, $M$ is the mass of the central object, and $c$ is the speed of light.
The geodesic equation is 
\begin{equation}
	\frac{d^2x^\mu}{d\tau^2} + \Gamma^\mu_{\nu\lambda} \frac{dx^\nu}{d\tau}\frac{dx^\lambda}{d\tau} = 0\,,
\end{equation}
where $\mu$, $\nu$, $\lambda$ are indices running from 0 to 3, $\tau$ is the proper time, and $\Gamma^\mu_{\nu\lambda}$ are the Christoffel symbols.
Restricting the motion to the plane $\theta=\frac{\pi}{2}$, we obtain for the radial acceleration:
\begin{equation}
\label{eq:rtau}
\begin{split}
	\frac{d^2 r}{d\tau^2} = - \frac{m}{\left(1 - \frac{2m}{r}\right) r^2}&\left(c^2 E_{\infty}^2-\left(\frac{dr}{d\tau}\right)^2\right)    \\
	& \quad +(r - 2m)\left(\frac{d\phi}{d\tau}\right)^2\,,
\end{split}
\end{equation}
where $E_{\infty}$ is the particle energy relative to its rest energy at $r=\infty$ ($dt/d\tau = E_{\infty}/(1-2m/r)$). 
Note that $E_{\infty}$ satisfies $E_{\infty}\geq 1$ for an unbounded particle (i.e. one that can reach infinity).
Looking at Eq.~\eqref{eq:rtau}, we can conclude that if $E_{\infty}\geq 1$, then the term
$$
c^2 E_{\infty}^2-\left(\frac{dr}{d\tau}\right)^2
$$
cannot be negative for $\left(\frac{dr}{d\tau}\right)^2<c^2$. 
If then we consider radial motion and $E_{\infty}\geq 1$, this means that for an observer moving with the particle, gravity cannot be repulsive. 
However, the contribution to the acceleration due exclusively to the velocity $dr/d\tau$ is repulsive.

The situation is slightly different if we consider an observer at infinity, whose measuring instruments are unaffected by gravity. 
The proper time of this observer is simply $t$.
Using $dt/d\tau = E_{\infty}/(1-2m/r)$ in Eq.~\eqref{eq:rtau}, we obtain that the radial acceleration measured by this observer is given by:
\begin{equation}
	\label{eq:rt}
\begin{split}
	\frac{d^2 r}{dt^2} &= -\frac{m}{\left(1-\frac{2m}{r}\right)r^2}\left(c^2 - 3 \left(\frac{dr}{dt}\right)^2\right)\\
	& +\frac{4m^2}{r^3} \frac{1 - \frac{m}{r}}{1-\frac{2m}{r}}+ (r - 2m) \left(\frac{d\phi}{dt}\right)^2 \,.
\end{split}
\end{equation}
Note here the dependence on the coordinate velocity $|dr/dt|$. In particular, for radial motion, if $|dr/dt|>c/\sqrt{3}$, the gravitational acceleration becomes repulsive at first order in $m$. 
Note that this is independent of the distance $r$: the fact that there is a repulsion only depends on the radial velocity.
Despite this effect appearing in general circular motion, in this paper we will focus on radial motion, and so we can simplify the above expression.
We can write Eq.~\eqref{eq:rt} in pure radial motion to first order in $m$ as
\begin{equation}
\label{eq:gv}
\frac{d^2 r}{dt^2} = -\frac{GM}{r^2}\left(1 -  \frac{3v^2}{c^2}\right) \equiv -g_v\,,
\end{equation}
where we defined $v=dr/dt$, put back the factor $GM$ and defined $g_v$ to make more explicit the changes relative to the standard $g \equiv GM/r^2$.
Note that if we consider a set of particles whose radial velocities are proportional to the radius $r$, say $v = H_0 r$, then we would obtain:
\begin{equation}
\frac{d^2 r}{dt^2} = -\frac{GM}{r^2} + \frac{3GM H_0^2}{c^2}\,,
\end{equation}
that is, there would be a constant gravitational repulsive contribution. 
More about this in Sec.~\ref{sec:ball}.

To check that this repulsion is not peculiar to Schwarzschild, let us know see how we can get a similar repulsion in the more general Parametrized Post-Newtonian (PPN) formalism.

\subsection{Repulsion in the PPN formalism}

We can also find a repulsive term in the PPN formalism, when considering the acceleration felt by two masses $m_1$ and $m_2$ at the 1PN order.
In global coordinates, as shown in~\cite{Damour1985}, that acceleration is given by
\begin{align}
\begin{split}
&\frac{d^2 \vec{r}}{dt^2} = - \frac{GM}{r^2} \vec{n} +\\
& \frac{	GM}{c^2 r^2} \left[\vec{n}
\left(
\frac{GM}{r}(4+2\nu) - v^2(1+3\nu) \right.\right.\\
&\left.\left. + \frac{3}{2}\nu (\vec{n}\cdot \vec{v})^2
\right) + (4-2\nu)(\vec{n}\cdot \vec{v}) \vec{v}
\right]\,,
\end{split} 
\end{align}
where $M=m_1+m_2$, $\vec{r}$ is the position vector between the two masses, $\vec{v}=d\vec{r}/dt$, $\vec{n} \equiv \vec{r}/r$, and $\nu = \mu/M$, $\mu$ being the reduced mass $\mu\equiv \frac{m_1 m_2}{m_1+m_2}$.
We can analyze purely radial motion by considering $\vec{v} = v \vec{n}$.
We obtain
\begin{align}
	\begin{split}
		&\frac{d^2 \vec{r}}{dt^2} = - \frac{GM}{r^2} \vec{n}\left[1-\left(3-\frac{7}{2}\nu\right)\frac{v^2}{c^2}\right] \\
		& + \frac{(GM)^2}{c^2 r^3}\vec{n}
		\left(
		4+2\nu\right)\,.
	\end{split} 
\end{align}
We notice that there is a repulsive term to first order in $GM$, given by
\begin{equation}
	g_{\rm rep} = \frac{GM v^2}{c^2 r^2} \left(3-\frac{7}{2}\nu\right)\,.
\end{equation}
So, the repulsion would depend on the value of $\nu$: a higher $\nu$ implies a lower repulsion.
However, the effect is always present and cannot be canceled. 
In fact, for any pair of masses $m_1$ and $m_2$, we have $0\leq \nu \leq 1/4$, where $\nu=1/4$ happens when the two masses are equal.
This means that in the scenario where repulsion is minimized, one still has
\begin{equation}
	g_{\rm rep} = 2.125 \frac{GM v^2}{c^2 r^2} > 0\,.
\end{equation}
Note that the Schwarzschild situation discussed in the previous section is a particular case where one mass vanishes, so $\nu=0$.
We can also observe in this situation that if the velocity is proportional to the radial distance there is an effective constant repulsive contribution, as we saw in the previous section. 
This repulsion can mimic a cosmological constant induced effect, as we will discuss in Sec.~\ref{sec:cosmology}.

\section{Expanding ball of dust embedded in Schwarzschild spacetime}
\label{sec:ball}

So far we have analyzed this gravitational repulsion effect in the Schwarzschild spacetime and in the more general PPN formalism.
We now want to find it in a cosmological-like situation.
For that, let us consider the Newtonian analogy to FLRW cosmology in a relativistic framing.
The analogy consists in considering an expanding uniform ball embedded in an exterior Schwarzschild spacetime.
This was considered in~\cite{Faraoni2020} to study an expanding ball of dust in GR.
Outside the expanding ball Birkoff's theorem enforces that the geometry is Schwarzschild. 
Inside the ball, we can use an FLRW metric, as is done in the Oppenheimer-Snyder dust collapse~\cite{Oppenheimer1939}.
Note that, despite having been originally developed to study the collapse of a dust ball, the same coordinate mapping is valid to study the ball's expansion. This was also done in~\cite{Faraoni2020}, and in this section we reproduce the relevant results.
We then choose two different sets of coordinates to describe this system. 
The metric in the interior of the expanding ball is given by
\begin{equation}
	ds^2 = -c^2d\bar{t}^2 + a^2(\bar{t}) \left(\frac{d\bar{r}^2}{1-K \bar{r}^2} + \bar{r}^2 d\Omega_2^2\right)\,,
\end{equation}
where $\bar{t}$ is the proper time, $\bar{r}$ is the co-moving radial coordinate, $K$ is the curvature index (that can have any value), and $d\Omega_2^2 = d\theta^2 + \sin^2 \theta \,d\phi$ is the solid angle element. 
The ball has uniform density $\rho = \rho(\bar{t})$.
Outside the ball, the metric is Schwarzschild by Birkhoff's theorem. 
We write it as
\begin{equation}
	ds^2 = -\left(1-\frac{2 m}{r}\right)c^2 dt^2 +  \frac{dr^2}{1-\frac{2m}{r}} + r^2 d\Omega_2^2\,.
\end{equation}
A coordinate transformation between interior and exterior can be found, yielding a simple expression for the radius:
\begin{equation}
	r = a(\bar{t}) \bar{r}\,,
\end{equation}
where the scale parameter $a(\bar{t})$ is positive,
and a more complicated expression for the time coordinate (that is not relevant for the discussion).
We can impose the Darmois-Israel junction conditions (see~\cite{Faraoni2020}), obtaining for geodesic motion
$\bar{r} = \bar{r}_0$, that is $\bar{r}$ is constant. 
The expanding ball's radius in Schwarzschild coordinates is then
\begin{equation}
	r(t) = a(\bar{t}) \bar{r}_0\,.
\end{equation}
We interpret the value of $r(t)$ to be of the order of the observable Universe radius.
The junction conditions also imply that 
\begin{equation}
	\begin{split}
&m = \frac{4 \pi \rho(\bar{t}) r(t)^3}{3} \\
&\Leftrightarrow \rho(t) r(t)^3 = \rho_0 r_0^3 = \frac{3m}{4 \pi} = \textrm{ const.} \,,
	\end{split}
\end{equation}
where a subscript $_0$ means that the quantities are evaluated at a specific time $t_0$.
Since there is no pressure, the surface of the ball follows geodesic motion in the exterior Schwarzschild spacetime.
Therefore, we can apply Eq.~\eqref{eq:rt} to analyze the expansion of the ball for an observer at infinity in Schwarzschild.
Since the surface of the ball follows a purely radial motion, we have
\begin{equation}
	\begin{split}
		\frac{d^2 r}{dt^2} &= -\frac{m}{\left(1-\frac{2m}{r}\right)r^2}\left(c^2 - 3 \left(\frac{dr}{dt}\right)^2\right)\\
		& +\frac{4m^2}{r^3} \frac{1 - \frac{m}{r}}{1-\frac{2m}{r}} \,.
	\end{split}
\end{equation}
Now, assuming that the expanding ball has a radius much larger than its Schwarzschild radius, we can neglect terms of order $o(m^2/r^2)$, and we get
\begin{equation}
	\frac{d^2 r}{dt^2} = -\frac{GM}{r^2}\left(1 - 3 \frac{\left(dr/dt\right)^2}{c^2}\right) \,.
\end{equation}
Replacing $r = a(t) \bar{r}_0$, we obtain
\begin{equation}
	\bar{r}_0\ddot{a} = -\frac{GM}{	\bar{r}_0^2 a^2}\left(1 - 3 \frac{\dot{a}^2 \bar{r}_0^2}{c^2}\right) \,,
\end{equation}
where here a dot means differentiation with respect to $t$.
To write this equation in a form more similar to a Friedmann equation, we plug $M =  \frac{4 \pi \rho_0 r_0^3}{3}$, obtaining
\begin{equation}
\label{eq:expandingball}
	\begin{split}
	\frac{\ddot{a}}{a} &= -\frac{4 \pi G \rho_0 a_0^3}{3 a^3}\left(1 - 3 \frac{\dot{a}^2 \bar{r}_0^2}{c^2}\right) \\
	&= -\frac{4 \pi G \rho_0 a_0^3}{3 a^3} + \frac{4 \pi G \rho_0 a_0^3 \dot{a}^2 \bar{r}_0^2}{c^2 a^3} \\
	&= -\frac{4 \pi G \rho(a)}{3} + \frac{4 \pi G \rho(a) r^2}{c^2} H^2\,,
	\end{split} 
\end{equation}
where we defined the Hubble parameter $H \equiv \dot{a}/a$, and here $r$ is the radius of the ball in Schwarzschild coordinates (changing in time).
Interestingly, Eq.~\eqref{eq:expandingball} reminds us of a Friedmann equation for a dust universe with a cosmological constant:
\begin{equation}
	\label{eq:friedmann}
\frac{\ddot{a}}{a} = -\frac{4 \pi G \rho(a)}{3} + \frac{\Lambda c^2}{3}\,.
\end{equation}
It is notable that, for an observer at infinity, the gravitational repulsion mimics the effect of a cosmological constant. 
In the next section we calculate which specific cosmological constant value is predicted by this simple model, and how a universe described by this expanding ball in general compares with $\Lambda$CDM cosmology.

\section{Application to cosmology: cosmological acceleration}
\label{sec:cosmology}

To translate the gravitational repulsion into a cosmological constant value, it is convenient to work with the (time-dependent) density parameters $\Omega_m \equiv \rho_m/\rho_c$ and $\Omega_\Lambda = \rho_\Lambda/\rho_c$, where $\rho_m$ is the matter density, $\rho_c\equiv \frac{3 H^2}{8\pi G}$ is the Friedmann critical density (in general varying with time), and $\rho_\Lambda =  \Lambda c^2/8\pi G$ is the energy density associated with the cosmological constant.
Then, assuming that the ball's density is $\rho_m$, we can write Eq.~\eqref{eq:friedmann} as
\begin{equation}
	\label{eq:newfriedmann}
	\begin{split}
	\frac{\ddot{a}}{a} &= -\frac{4 \pi G}{3} \rho_m + \frac{8\pi G}{3} \rho_\Lambda \\
	\frac{\ddot{a}}{a} &= -\frac{4 \pi G}{3} \rho_m + \frac{8\pi G}{3} \frac{\Omega_\Lambda}{\Omega_m}\rho_m\,.
	\end{split}
\end{equation}
Then, comparing the r.h.s. of Eqs.~\eqref{eq:expandingball} and \eqref{eq:newfriedmann}, we get the equivalence:
\begin{equation}
\label{eq:lambdaequiv}
\frac{r^2}{c^2} H^2 =  \frac{2}{3} \frac{\Omega_\Lambda}{\Omega_m}\,.
\end{equation}
Taking values at the present time $t_0$, we can write $r = r_0 = \alpha c/H_0$, where $\alpha$ is of order 1, we obtain (at $t_0$):
\begin{equation}
	\label{eq:omega}
	\Omega_\Lambda =  \frac{3}{2}\alpha^2 \Omega_m\,.
\end{equation}
Eq.~\eqref{eq:omega} is remarkable because it shows that $\Omega_\Lambda$ is of the same order as $\Omega_m$ in this expanding ball model.
So, if the observer at infinity could measure the matter density of the sphere, and if that matter density were similar to our universe's matter density, then he would infer the existence of a cosmological constant with a value very close to ours. 
This observer could then be surprised because that value would correspond to a dark energy density of the same order of the matter density, calling it a coincidence problem. 
However, in that case, there was never a problem, because this was a necessary consequence of general relativity.
The interest of this result also comes from the fact that the so-called ``Newtonian cosmology'' is exactly this system of a uniform expanding ball, but treated classically.
Thus treating this Newtonian picture in a general relativistic framework effectively yields a effective cosmological constant.

We now are tempted to apply this result to our own universe. However, it is not straightforward. 
First, we live inside the universe, so we could not be an observer at infinity.
Second, this model has special locations (for example the center of the sphere), and so violates the Copernican principle. Indeed, if our Universe is governed by the Copernican principle, it is non-trivial to even define what an observer at infinity is.

We can nevertheless question how we could quantify the effect of a possible gravitational repulsive term across many pairs of galaxies over the whole universe. 
Would it effectively yield something like a cosmological constant?

One way to approach the problem is considering derivatives relative to proper time $\tau$ instead of coordinate time $t$.
In that case, the same procedure applies, and we use Eq.~\eqref{eq:rtau} instead of Eq.~\eqref{eq:rt}.
For radial motion, and assuming that the spherical ball's radius is much larger than its Schwarzschild radius ()neglecting terms that are $o(m^2/r^2)$), Eq.~\eqref{eq:rtau} becomes
\begin{equation}
\label{eq:geodesicproper}
	\frac{d^2 r}{d\tau^2} = -\frac{GM}{r^2}\left(E_{\infty}^2 - \frac{\left(dr/d\tau\right)^2}{c^2}\right) \,.
\end{equation}
For an expanding dust ball, we can replace $r = a(\tau) \bar{r}_0$, obtaining (here a dot means differentiation with respect to proper time):
\begin{equation}
	\bar{r}_0 \ddot{a} = -\frac{GM}{a^2 \bar{r}_0^2}\left(E_{\infty}^2 - \frac{\dot{a}^2  \bar{r}_0^2}{c^2}\right) \,,
\end{equation}
which simplifies to
\begin{equation}
	\frac{\ddot{a}}{a} = -\frac{GM}{a^3 \bar{r}_0^3}\left(E_{\infty}^2 - H^2 \frac{r^2}{c^2}\right) \,,
\end{equation}
where we defined $H \equiv \dot{a}/a$.
Finally, replacing $M=4\pi \rho_0 r_0^3/3$, we obtain
\begin{equation}
	\frac{\ddot{a}}{a} = -\frac{4\pi G \rho(a)}{3}\left(E_{\infty}^2 - H^2 \frac{r^2}{c^2}\right)\,,
\end{equation}
so we get a very similar result as for an observer at infinity.
In order for the first term to match the Friedmann equation term, let us set $E_\infty = 1$ (this just means that when $r\to\infty$, $dt = d\tau$).
Then, comparing with Eq.~\eqref{eq:newfriedmann}, we obtain
\begin{equation}
	\frac{r^2}{c^2} H^2 =  \frac{2\Omega_\Lambda}{\Omega_m}\,.
\end{equation}
Following the same procedure as before, we plug in $r=r_0=\alpha c/H_0$ and $H=H_0$, obtaining
\begin{equation}
\Omega_\Lambda = \frac{1}{2}\alpha^2 \Omega_m \,.
\end{equation}
So an observer co-moving with the ball would measure an $\Omega_\Lambda$ very closely related to $\Omega_m$.

It could be pointed out that when using the proper time another constraint is valid from the definition of the line element, namely (for radial motion and massive particles)
\begin{equation}
\label{eq:lineconstraint}
	1 = \left(1 - \frac{2m}{r}\right) \left(\frac{dt}{d\tau}\right)^2 - \left(1 - \frac{2m}{r}\right)^{-1} \left(\frac{dr}{c \,d\tau}\right)^2\,,
\end{equation} 
which would have given $dr/cd\tau$ in terms of $E_\infty$.
We could then analytically remove the acceleration's dependence on the velocity $dr/d\tau$. 
However, in this particular case we are more focused in analyzing the effects coming the gravitational field itself, which is explicitly encoded in the geodesic equation.
To better understand this, let us consider the equation of motion in GR for a charged particle with mass $m_p$ and charge $q$.
It is given by
\begin{equation}
	m_p\frac{d^2x^\mu}{d\tau^2} + m_p\Gamma^\mu_{\nu\lambda} \frac{dx^\nu}{d\tau}\frac{dx^\lambda}{d\tau} = q F^{\mu}{}_{\nu}\frac{dx^\nu}{d\tau}\,,
\end{equation}
where $F^{\mu}{}_{\nu}$ is the electromagnetic field tensor.
The interpretation of this equation is that, as $q F^{\mu}{}_{\nu}\frac{dx^\nu}{d\tau}$ represents the Lorentz force, $-m_p\Gamma^\mu_{\nu\lambda} \frac{dx^\nu}{d\tau}\frac{dx^\lambda}{d\tau}$ represents the gravitational force.
So one should focus on that specific term in order to analyze gravitational repulsion. 
Moreover, Eq.~\eqref{eq:geodesicproper} is also valid for a massless particle, whereas Eq.~\eqref{eq:lineconstraint} is only valid for massive test particles.
So, the more general equation to use is indeed Eq.~\eqref{eq:geodesicproper}, which yields a cosmological constant like repulsion in the Newtonian cosmology limit.

So, in this Newtonian cosmological model analyzed in a general relativistic framework, there is an emergent cosmological constant that is naturally of the same order as $\Omega_m$.
However, for $E_\infty \geq 1$, net gravitational repulsion cannot be achieved in this case.

\subsection{Dynamical analysis}

Having shown that a general relativity enhanced Newtonian cosmology can mimic a cosmological constant, it is of interest to analyze its differences relative to standard $\Lambda$CDM cosmology where the universe is described by a Friedmann-Lema\^{i}tre-Robertson-Walker (FLRW) model.
In our Newtonian cosmology, we consider an observer at infinity looking at an expanding ball of pressureless dust (the ``universe'').
So, for this observer, Eq.~\eqref{eq:expandingball} describes the scale factor's evolution.
Note that in this situation there is no equivalent to the standard Friedmann equation given by
\begin{equation}
	H^2 = \frac{8\pi G}{3} \rho - \frac{k c^2}{a^2}\,,
\end{equation}
where $H$ is the Hubble parameter and $k=\{-1,0,1\}$ indicates the spatial curvature of the FLRW model. 
Indeed, Eq.~\eqref{eq:expandingball} corresponds in standard cosmology to only the Friedmann acceleration equation given by Eq.~\eqref{eq:friedmann} for pressureless dust.

Let us then analyze Eq.~\eqref{eq:expandingball} subject to the constraints $H(t=t_0) = H_0>0$ and $a(t=t_0)=a_0>0$.
We notice first that there are 3 regimes, depending on the value of $\dot{a}$:
\begin{enumerate}
	\item if $\dot{a}^2=c^2/3 r_0^2$, then $\ddot{a} = 0$ and so we just have 
	\begin{equation}
		a(t) = \frac{c t}{\sqrt{3} r_0}\,.
	\end{equation}
This evolution is similar to a Milne universe for a FLRW metric.
	\item if $\dot{a}^2>c^2/3 r_0^2$, we can analyze how the dynamics changes if $\dot{a}^2 = c^2/3 r_0^2(1+\epsilon)$ (i.e. if one deviates slightly from the previous situation).
	Then, $\ddot{a}>0$ and Eq.~\eqref{eq:expandingball} becomes for $\epsilon \ll 1$
	\begin{equation}
		\label{eq:epsilon}
		\begin{split}
		 \frac{c}{2\sqrt{3} r_0}\dot{\epsilon} &= \frac{4 \pi G \rho_0 a_0^3}{3 a^2} \epsilon \,,
		\end{split} 
	\end{equation}
	meaning that $\epsilon$ grows (approximately) exponentially, making the configuration $\epsilon=0$ unstable.
	So, if $\dot{a}^2>c^2/3 r_0^2$ at some time, and $a>0$, the inequality will hold for the whole motion, and so $\ddot{a}>0$ at all times.
	
	\item if $\dot{a}^2<c^2/3 r_0^2$, then $\ddot{a}<0$, and we can analyze how the dynamics changes if $\dot{a}^2 = c^2/3 r_0^2(1-\epsilon)$.
	This reduces to the previous situation for negative $\epsilon$, so Eq.~\eqref{eq:epsilon} holds.
	In principle, there is nothing stopping $\dot{a}$ to vanish and change sign in this situation.
	However, if $\dot{a}$ becomes negative, $a(t)$ starts decreasing and could eventually become $0$, so we could get a ``Big Crunch''.
\end{enumerate}

So, in a (Newtonian) universe where $\ddot{a}>0$, one has  $\dot{a}^2>c^2/3 r_0^2$ at all times.

\subsection{Comparison with $\Lambda$CDM}

In this section, we compare how such a Newtonian universe, as measured by an observer whose instruments are unaffected by gravity, compares with a standard $\Lambda$CDM.
For that, we consider a flat universe with matter and a cosmological constant, and that the cosmological constant like terms match today ($t=t_0$) in the Friedmann equation and in Eq.~\eqref{eq:expandingball}.
In particular, this means that Eq.~\eqref{eq:lambdaequiv} is satisfied at $t=t_0$.
We can compare with the evolution of the standard $\Lambda$CDM parameter $\Omega_{\Lambda}(a)$ by defining a ``Newtonian'' equivalent defined by
\begin{equation}
\label{eq:omegaLN}
	\Omega_\Lambda^N = \frac{3 r^2 H^2}{2 c^2} \Omega_m = \frac{4 \pi G r^2 \rho_m}{c^2}\,,
\end{equation}  
where we used $\Omega_m = 8 \pi G\rho_m/3 H^2$.
Then, since $r^2 \propto a^2$ and $\rho_m \propto a^{-3}$, the evolution of $\Omega_\Lambda^N$ with the scale factor is then simply given by $\Omega_\Lambda^N \propto a^{-1}$.

\begin{figure}
\centering
\includegraphics[width=\columnwidth]{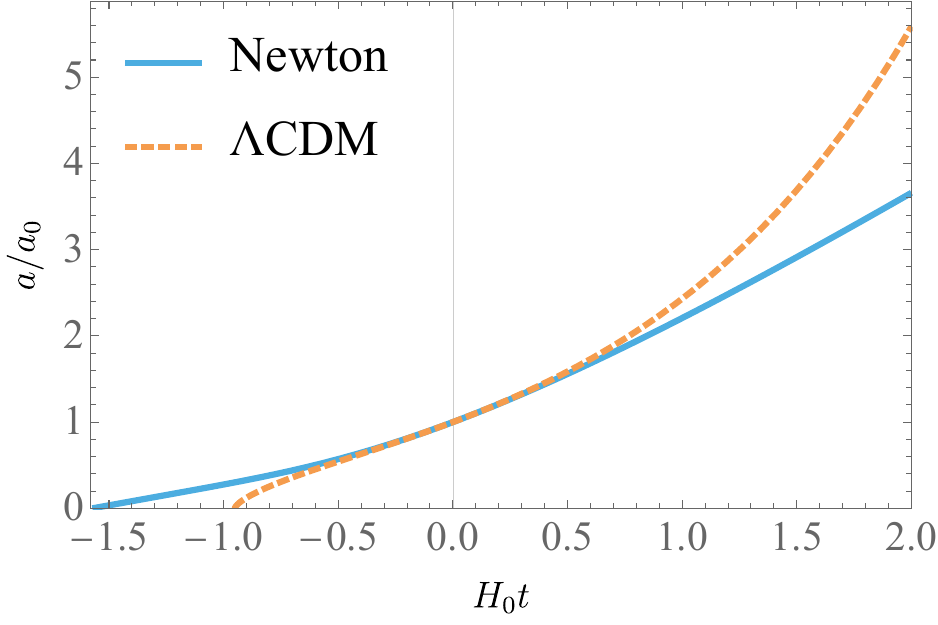}
\caption{
\label{fig:scaleplot}
Scale factor evolution as a function of time for an expanding Newtonian ball of dust and for standard $\Lambda$CDM by numerically solving Eqs.~\eqref{eq:simpleNewton} and \eqref{eq:simpleCDM}. 
At $t=0$ (today), $a=a_0=1$ and the values of $\Omega_{\Lambda,0}$ and $\Omega_{m,0}$ match for the two curves. 
The plots are up to $H_0 t = 2$ in the future and stop when $a=10^{-3}$ in the past.
More time elapsed since $a=10^{-3}$ in the Newtonian model than in $\Lambda$CDM.
In the future, $\Lambda$CDM expands faster than the Newtonian model. 
}
\end{figure}

On the other hand, the $\Lambda$CDM dependence for $\Omega_\Lambda$ can be obtained by using the definition $\Omega_\Lambda = 8 \pi G\rho_\Lambda/3 H^2$, and using $H^2/H_0^2=(1-\Omega_{\Lambda,0})a_0^3/a^3 + \Omega_{\Lambda,0}$ and $\rho_\Lambda = 3 \Omega_{\Lambda,0}/8\pi G H_0^2$ (note that this is constant). We then obtain:
\begin{equation}
\Omega_\Lambda = \frac{\Omega_{\Lambda,0}}{(1-\Omega_{\Lambda,0})a_0^3/a^3 + \Omega_{\Lambda,0}}\,.
\end{equation}
To compare the evolution in time of $\Omega_\Lambda$ and $\Omega_\Lambda^N$, we can look at large timescales by numerically solving the differential equations corresponding to the Newtonian (Eq.~\eqref{eq:expandingball}) and the $\Lambda$CDM universes (Eq.~\eqref{eq:friedmann}), which can be simplified respectively to:
\begin{align}
\label{eq:simpleNewton}
\text{Newton: } a'' &= -\frac{\Omega_{m,0}}{2a^2} + \left(\frac{a'}{a}\right)^2 \Omega_{\Lambda,0}\,,\\
\label{eq:simpleCDM}
\text{$\Lambda$CDM: } a'' &= -\frac{\Omega_{m,0}}{2a^2} + a \,\Omega_{\Lambda,0}\,,
\end{align}
where a prime means differentiation with respect to $H_0 t$, and we set $a_0=1$. 
Note that in both systems the density parameters measured at $t=t_0$ are the same.
In Fig.~\ref{fig:scaleplot}, we plot the scale factor as a function of time for $\Omega_{\Lambda,0}=0.685$ and $\Omega_{m,0} = 0.315$.
The plot starts when $a=10^{-3}$ and stops at $H_0 t=2$. 
Today is $t=0$.
We observe that more time elapsed since $a=10^{-3}$ in the Newtonian model than in $\Lambda$CDM.
In the future, $\Lambda$CDM expands faster than the Newtonian model. 

\begin{figure}
	\centering
	\includegraphics[width=\columnwidth]{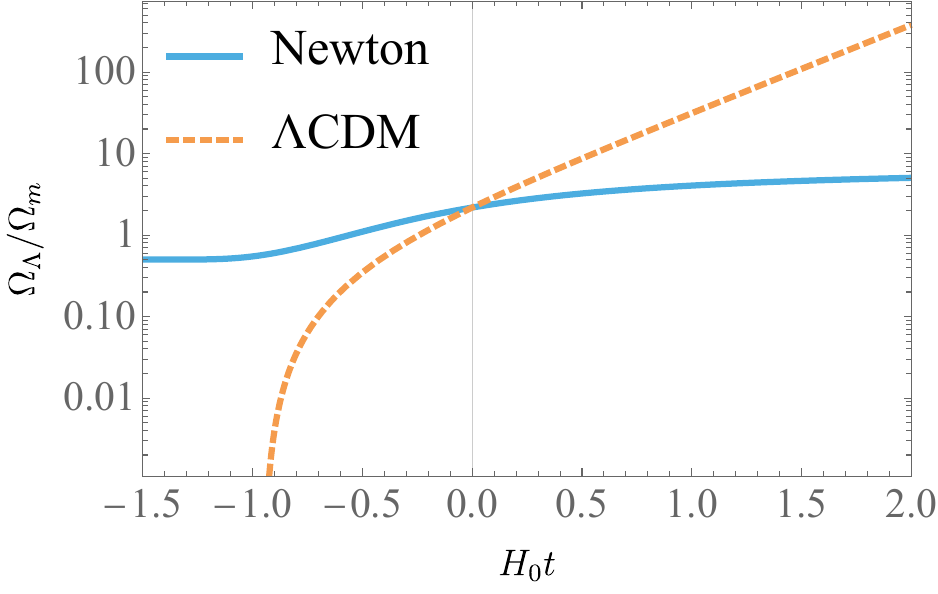}
	\caption{
		\label{fig:omegaratio}
		Evolution of the ratio $\Omega_\Lambda/\Omega_m$ as a function of time for an expanding Newtonian ball of dust and for standard $\Lambda$CDM. 
		At $t=0$ (today), $a=a_0=1$ and the value of the ratio $\Omega_{\Lambda,0}/\Omega_{m,0}$ matches for the two curves. 
		The plots are up to $H_0 t = 2$ in the future and stop when $a=10^{-3}$ in the past.
		In the Newtonian model, the ratio is always of order 1, whereas in $\Lambda$CDM it spans many orders of magnitude.
	}
\end{figure}

Another useful comparison is the ratio $\Omega_\Lambda/\Omega_m$ as a function of time. 
Indeed, it has been argued that $\Lambda$CDM assumes that we are in a special time where $\Omega_\Lambda/\Omega_m$ is of order 1, when it could have in principle any value between 0 (in the distant past) and $\infty$ (in the distant future).
The fact that today the energy density attributed to dark energy is of the same order as the matter energy density is the so-called coincidence problem.
In Fig.~\ref{fig:omegaratio}, we plot $\Omega_\Lambda/\Omega_m$ for the Newtonian model (using Eq.~\eqref{eq:omegaLN}) and for $\Lambda$CDM (simply given by $a^3\Omega_{\Lambda,0}/\Omega_{m,0}$). 
We observe that for the whole history of the Universe, $\Omega_\Lambda^N/\Omega_m$ is of order 1 in the Newtonian model, so the model is not affected by the coincidence problem.

\section{Discussion and Conclusion}
\label{sec:conclusion}

In this work, we started by reviewing situations in which gravitational repulsion can happen in general relativity. 
Despite being relatively straightforward to derive, this effect is rarely discussed in the literature.
This repulsion clearly exists for an observer at infinity, and depends on the square of the radial velocity of the object.

We then analyzed the system typically associated with Newtonian cosmology: an expanding ball of dust. 
By studying this system in a general relativistic framework, we found that in the Newtonian approximation where $m/r \ll 1$ there is an effective cosmic repulsive acceleration for an observer at infinity.
Then, we compared this Newtonian cosmology with a $\Lambda$CDM cosmology if we had the same cosmological parameters today. 
We found that the Newtonian cosmology provides an older Universe, and that it does not suffer from the cosmic coincidence problem.

Let us make clear that this is only valid for an observer at infinity, unaffected by gravity. 
For an observer co-moving with the uniform, expanding ball, if there are no other forces besides gravity, no cosmic repulsion is observed.

This work then clearly suggests one way to overcome some problems in cosmology today. 
If by some mechanism the late-time universe dynamics could be described such that the radial acceleration were given by Eq.~\eqref{eq:gv} or similar, then we would have an explanation for cosmic acceleration that would just involve gravity, simultaneously solving the cosmic coincidence problem.
Alternatively, this is also valid if the cosmological limit of some theory of gravitation is similar to Eq.~\eqref{eq:expandingball}.
These research avenues will be the subject of future work.

Moreover, it would be of interest to investigate the effect of this gravitational repulsion in relativistic astrophysical jets (e.g. blazar jets \cite{Blandford2019}), since in that case the velocities are relativistic and the Earth could indeed be modeled as an observer at infinity. 
This interesting topic is also left for future work.

\section*{ACKNOWLEDGMENTS}

The author would like to thank Bob Wagoner for helpful comments and discussions, and KIPAC for financial support. 

\appendix

\bibliography{RepulsiveGravity}

\end{document}